\documentclass[useAMS,usenatbib]{mn2e}
\usepackage{graphicx}
\usepackage[latin1]{inputenc}
\usepackage{color}
\usepackage{times}
\usepackage{natbib}
\usepackage{setspace}
\usepackage{amsmath}
\usepackage{subfigure}
\usepackage{rotating}
\usepackage{lscape}

\newif\ifAMStwofonts
\AMStwofontstrue
\definecolor{red}{rgb}{1,0.,0.}

\title[The PM2GC] {The Padova-Millennium Galaxy and Group Catalogue (PM2GC): the group-finding method and the PM2GC catalogues of group, binary and single field galaxies}

\author
[Calvi et al.]{Rosa Calvi$^{1}$\thanks{E-mail: rosa.calvi@unipd.it},
  Bianca M. Poggianti$^{2}$, Benedetta Vulcani$^{1,2}$\\
$^{1}$Astronomical Department, Padova University, Italy\\
$^{2}$INAF-Astronomical Observatory of Padova, Italy\\}

\begin{document}

\date{Accepted .... Received ..}

\pagerange{\pageref{firstpage}--\pageref{lastpage}} \pubyear{2010}

\maketitle
\label{firstpage}
\begin{abstract}
We present the construction and describe the properties of the
Padova-Millennium Galaxy and Group Catalogue (PM2GC), a galaxy
catalogue representative of the general field population in the
local Universe. We characterize galaxy environments by identifying
galaxy groups at 0.04$\leq$z$\leq$0.1 with a Friends-of-Friends
(FoF) algorithm using a complete sample of 3210  galaxies brighter
than $M_B=-18.7$ taken from the Millennium Galaxy Catalogue (MGC,
\citet{b26}), a 38$\rm deg^2$ photometric and spectroscopic
equatorial survey. We identified 176 groups with at least three
members, comprising in total 1057 galaxies and representing
$\sim$43 per cent of the general field population in that redshift
range. The median redshift and velocity dispersion of our groups
are 0.0823 and 192 $\rm km \, s^{-1}$, respectively. 88 per cent
of the groups have fewer than ten members, and 63 per cent have
fewer than five members. Non-group galaxies were subdivided into
``binary'' systems of two bright close companions, and ``single''
galaxies with no companion, in order to identify different
environments useful for future scientific analysis. We performed a
detailed comparison with the 2PIGG catalog to validate the
effectiveness of our method and the robustness of our results.
Galaxy stellar masses are computed for all PM2GC galaxies, and
found to be in good agreement with Sloan Digital Survey Data
Release 7 (SDSS-DR7) mass estimates. The catalogues of PM2GC groups,
group properties and galaxy properties in all environments are
publicly available on the World Wide Web.
\end{abstract}


\section{Introduction}
\label{s:introduction}
In the $\Lambda$CDM model of the Universe, the growth of
large-scale structure occurs hierarchically with the most massive
structures, like clusters and superclusters, formed through the
accretion of smaller halos which continuously interact and merge,
acquiring infalling galaxies along filaments.

Historically, the major discoveries concerning the impact of
hierarchical assembly and thus environment on galaxy evolution
come from the studies conducted on clusters. The Butcher-Oemler
effect, the morphology vs. density relation and the star formation
vs. density relation have shown that the properties of galaxies
within clusters are strongly correlated with the local environment
and evolve with redshift (see, e.g., \citealt{b1,b2,b3,b4,b5}). In
agreement with the observed trends a presumable scenario is that
the clustering process itself would drive the evolution of the
galaxy properties and the typical increase of the early-type
fraction with decreasing redshift would be due to gas-rich,
star-forming disk galaxies
 which fall into clusters at higher redshift
(z$\sim$0.5-1) and have their gas reservoir depleted by some
mechanism that transforms them into red, bulge-dominated and
quiescent galaxies at z$\sim$0.

Among the numerous mechanisms proposed to deplete the reservoir of
gas in late-type galaxies are ram-pressure and/or tidal stripping
\citep{c5}, interaction with the cluster potential \citep{c6} and
repeated high-velocity encounters (''harassment'') \citep{c7}.
However, recent results \citep{b11,b16} have focussed on the
hypothesis that the anomalous fraction of early-type galaxies in
clusters is the consequence of some pre-process which takes place
in \textit{groups} before galaxies are accreted into the cluster.
This hypothesis is supported by the observations of a decline in
star formation rate (SFR) in the outskirts of clusters, well
outside the virial radius \citep{c3,c4}. Although \citet{ber}
found that the pre-processes in the group environment seem not to
produce a large effect, it is nevertheless true that the
mechanisms like galaxy mergers and strangulation that efficiently
act also in groups could play a very important role in the
formation of the galaxy populations, as it has been shown by
combining high-resolution N-body simulations with semi-analytic
models for galaxy evolution
\citep{c9,c10,c12,c13,c14,c15,c16,c17,c18,c19,c20,c21}.

Compared to clusters galaxy groups are more difficult to detect
because they have a lower density with respect to the background
galaxy population, but are much more common in the Universe. Today
over 50 per cent of galaxies are in these systems and span a wide
range in local density showing properties which range from
cluster-like to field-like \citep{b11}. For this reason they are
thought to be a key environment to investigate galaxy evolution
and to provide a clear framework on the nature of the physical
galaxy transformation mechanisms.

Until recently, the difficulties in obtaining large, unbiased
samples of groups have forced most of the studies to use small
samples selected, for example, from the Hickson compact group
catalogue \citep{b14}, from the CfA redshift survey
\citep{b12,b13}, and from X-ray surveys \citep{b15,b16}. Only with
the advent of large galaxy redshift surveys, such as the Two
degree Field Galaxy Redshift Survey (2dFGRS), the Sloan Digital
Sky Survey (SDSS) and the Canadian Network for Observational
Cosmology Redshift Survey (CNOC2), has it become possible to
generate large group catalogues in the local Universe (e.g.
\citealt{b9,b17,b18,b19,b20,b21,b22,b10,b23}), and at intermediate
redshift \citep{b24,c1}.

The methods used to identify galaxy groups depend on several
assumptions and there is no unique group-finder algorithm to
assign them members. The most popular is the Friends-of-Friends
(FoF) algorithm applied to galaxy redshift surveys, first proposed
by \citet{b9}. Other approaches, which strongly depend on the
properties of dark matter haloes, can be found in
\citet{b31,b32,b33,b34}. The 2PIGG (2dFGRS Percolation-Inferred
Galaxy Group) catalogue \citep{b10} and the SDSS group catalogue
\citep{c2} are two of the largest available samples of galaxy
groups which use ``realistic'' mock catalogues to calibrate the
parameters associated with the group-finder algorithm.

The motivation for another group catalogue and for our work is to
provide a new dataset characterized by both high spectroscopic
completeness, to define galaxy environment well, with high quality
imaging, to investigate galaxy properties such as galaxy
morphologies which could not explored in detail in other
catalogues.

In this paper we present the construction and describe the
properties of the Padova-Millennium Galaxy and Group Catalogue
(PM2GC), a database of groups and galaxies at low redshift fully
provided, easily upgradable and easily to be consulted. The PM2GC
redshift range (0.03$\leq$z$\leq$0.11)
is similar to WINGS (WIde-field Nearby Galaxy-cluster Survey), a
survey of 77 X-ray selected galaxy clusters \citep{b25}. The
combination of PM2GC and WINGS allows
the study of the properties and the evolution of galaxies in the
widest possible range of environments in the local Universe and to
understand the origin of the observed trends of galaxy properties
versus environment and the relation between galaxy star formation
histories and the growth history of structures.

This paper is structured as follows. In $\S$2 we present the
dataset (MGCz) used and give an overview of the survey; in $\S$3
we provide a description of our approach to identify groups and
our group-finder algorithm linking criteria; in $\S$4 we show the
properties of groups and galaxies in the different environments
identified; in $\S$5 we derive the galaxy stellar masses comparing
them with SDSS-DR7 masses; in $\S$6 we test the reliability of our
group sample comparing it with the 2dFGRS group catalogue (2PIGG,
Eke et al. 2004); in $\S$7 we present the publicly released PM2GC
catalogues and finally in $\S$8 we present a summary. Throughout
the paper we adopt $H_{0}=70 \, \rm km \, Mpc^{-1} s^{-1}$,
$h=H_{0}/$100, $\Omega_{m}=$0.3 and $\Omega_{\lambda}=$0.7. All
magnitudes are expressed in the Vega system unless otherwise
stated.

\section{The Data}

To build a catalogue that satisfies our requirements of
spectroscopic and photometric completeness, we used a set of
galaxies derived from the Millennium Galaxy Catalogue (MGC)
\citep{b26,c11,b29}, a B-band imaging survey, both deep and wide,
which provides a high quality, complete representation of the
nearby galaxy populations.

A detailed description of the survey strategy, the photometric and
astrometric calibration and the object detection and
classification can be found in \citet{b26}. In brief, the survey
extends along an equatorial strip covering an area of $\sim$37.5
deg$^{2}$ and consists of 144 overlapping fields taken with the
WFC four-CCD mosaic on the Isaac Newton Telescope, with a uniform
isophotal detection limit of 26.0 mag arcsec$^{-2}$. The catalogue
contains about one million of objects reduced by the Cambridge
Astronomy Survey Unit (CASU) \citep{b27} and classified using
Source Extractor (SE{\footnotesize XTRACTOR}, \citet{b28}). The
entire set of objects, spanning the range 16$\leq B_{MGC} <$24,
was next divided into two magnitude ranges to better address the
division between stars and galaxies: the MGC-BRIGHT catalogue,
which contains all objects with $B_{MGC}<$20 mag, and the
MGC-FAINT catalogue which contains the others.

For this paper we selected a sample of galaxies from the MGCz
catalogue - a version of the total MGC database available on DVD -
that is the spectroscopic extension of MGC-BRIGHT. It was built
upon the
redshifts provided by the 
2dFGRS and the SDSS-EDR/DR1, in which the MGC survey region is
fully contained, and completed with redshifts taken by the MGC
team at the Anglo Australian Telescope using the 2dF facility
\citep{b29}, as well as redshifts from the NASA Extragalactic
Database (NED), the 2dF QSO Redshift Survey (2QZ), the Paul
Francis' Quasar Survey and of some low surface brightness
galaxies. The total spectroscopic completeness of galaxies
obtained by MGC team is greater than 96 per cent for $B_{MGC}<$20,
so we have no need to apply a statistical completeness correction
to  sample\footnote{For the additional completeness test see \S
3}.

At the beginning, we extracted galaxies at 0.03$\leq$z$\leq$0.11;
we chose this redshift range to avoid galaxies too close by whose
spectra only sample the central regions, while remaining at
sufficiently low redshifts to retain a deep absolute magnitude
completeness limit. Absolute B-band magnitudes were obtained
k-correcting the observed SE{\small XTRACTOR} `BEST' magnitudes
(MAGAUTO, except in crowded region where the ISOCOR magnitude was
used instead), corrected for Galactic extinction. The
k-corrections were taken from \citet{b30}, using the galaxy
redshift and the Sloan galaxy color provided in the MGCz catalogue
(hereafter, MGC-SDSS).

\begin{figure}
    \centering
    \includegraphics[scale=0.4]{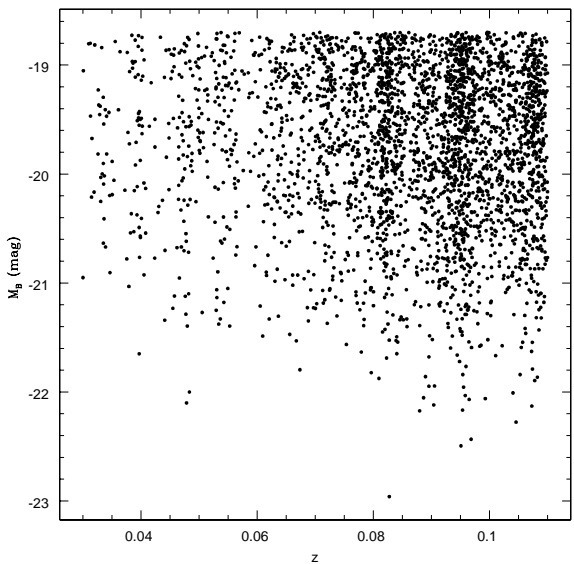}
    \caption{Absolute magnitude in B band vs. redshift for our data sample.} \label{F0}
\end{figure}

To build our catalogues we used only 3210 bright galaxies with a
magnitude $M_{B}<$-18.7 corresponding to the k-corrected
$B_{MGC}=$20 magnitude at our redshift upper limit. In
Fig.\ref{F0} a plot of absolute magnitude vs. redshift for this
sample is shown. The high spectroscopic completeness to
$B_{MGC}<$20, coupled with the
photometric depth, makes it a 
complete absolute magnitude-limited sample.
\begin{table}
\centering
\begin{tabular}{lcc}
\hline
\hline
\textbf{Environment} & \multicolumn{2}{c}{\textbf{Number of galaxies}}\\
            & $0.04\leq z\leq 0.1$ & $0.03\leq z\leq 0.11$\\
\hline
\hline
Groups & 1057 & $-$\\
field-single & 846 & 1141\\
field-binary & 367 & 490\\
Mix sample & 208 & 522\\
General field &  2460  & 3210\\
\hline
\end{tabular}
\centering
\caption{List of the number of galaxies in the different environments.}\label{t1}
\end{table}

\section{Group building method}
The approach we used to identify galaxy groups is similar to that
adopted by \citet{b38}, and is based on a plain FoF algorithm.
According to FoF criteria, two galaxies, \textit{i} and
\textit{j}, are physically related and join the same group if
their distances in the projected direction ($D$) and in the
line-of-sight ($V$) are less than some fixed thresholds, i.e.
\begin{equation}
    D_{ij}\leq D_{L};\quad V_{ij}\leq V_{L}
\end{equation}
\textit{D$_{L}$} and \textit{V$_{L}$} are called ''linking
lengths'' and link together all galaxies within a particular
linking volume.
We chose these lengths to take into account the typical
gravitational bounds of groups and to follow a similar approach of
that used to identify groups at high redshift in the ESO Distant
Cluster Survey \citep{c22,c23,c24,c25}. Similarly to other studies
\citep{b38,c1}, we adopted for a linking volume a cylinder centred
on each galaxy with radius
\begin{equation}
    {D_{L}}=0.5h^{-1} Mpc
\end{equation}
corresponding to a density contour $\frac {3}{4\pi D_{L}^{3} n} =
216.6 gal^{-1}$, being $n$ the mean observed number density of
galaxies in the total sample. The line-of-sight depth $V_{L}$ is
equal to three times the velocity dispersion, fixed at 500 $\rm km
\, s^{-1}$ rest frame, of the galaxy redshift.

For each galaxy in our sample brighter than $M_{B}=$-18.7, we
obtained its first neighbours in the cylinder defined above, and
added
to these, by a recursive procedure, neighbors of neighbors, until
no more are found. The resulting system we defined the ``trial''
group.
Only systems with at least three galaxies were further considered
as group candidates.

As a second step, we computed for each trial group its median
geometric centre, median redshift and velocity dispersion using
the statistical methods by \citet{b39}, considering the gapper
scale estimator for groups with less than ten galaxies and the
biweighted scale estimator for more populous systems.
A galaxy was considered member of a group if its spectroscopic
redshift lay within $\pm 3\sigma$ from the median group redshift
and if it was located within a projected distance of $\pm 1.5 \,
R_{200}$ from the geometrical centre. $R_{200}$ is an
approximation of the virial radius - the radius which delimites a
sphere with mean interior density 200 times the critical density -
computed as in \citet{b40}
\begin{equation}
    R_{200}=\frac{1.73\sigma}{1000 km s^{-1} \sqrt{\Omega_{\Lambda}+\Omega_{0}(1+z)^{3}}}h^{-1}Mpc
\end{equation}
where $\sigma$ and $z$ are the group's velocity dispersion and
median redshift, respectively.

We iterated the second step several times, recalculating every
time the group redshift, velocity dispersion and $R_{200}$,
 and moving to the next iteration only those groups with at least
three members. The process stops when the last two iterations have
identical output. At most, three iterations were sufficient to
reach convergence. We consider members of the final groups only
those galaxies that are within $1.5 \, R_{200}$ from the group
centre and 3$\sigma$ from the group redshift.

Using this method we obtained a sample of 176 galaxy groups in the
redshift range 0.04$\leq$z$\leq$0.1 containing in total 1057 group
members with magnitude $M_{B}<$-18.7.

Groups below z$\sim$0.04 and above z$\sim$0.1 are disregarded in
the following analysis, because, for a maximum velocity dispersion
of 800 $\rm km \, s^{-1}$, and due to the redshift limits of our
original sample ($0.03-0.11$), they can suffer from spectroscopic
incompleteness.
\begin{figure*}
    \centering
    \vspace{-350pt}
    \includegraphics[scale=0.9]{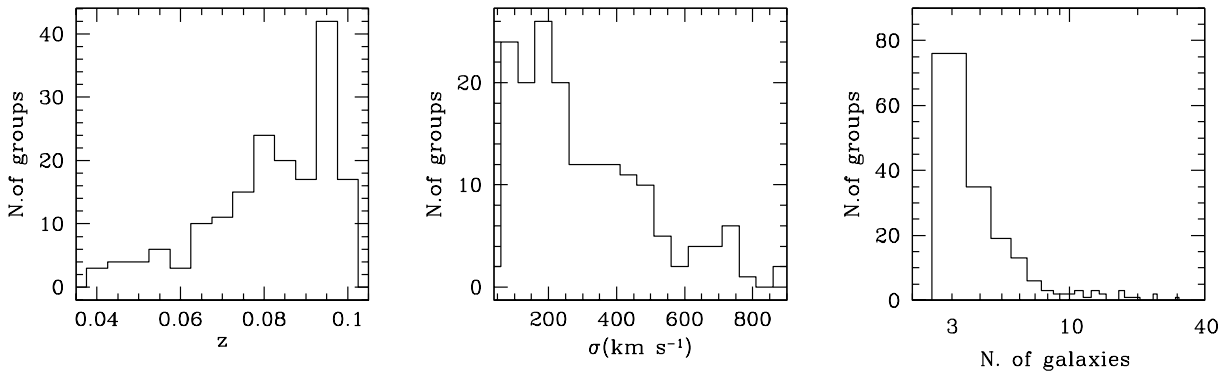}
    \caption{Redshift ({\it z}) distribution (left), velocity dispersion ($\sigma$) distribution (centre) and number of members distribution (right) of the 176 groups at 0.04$\leq$z$\leq$0.1.} \label{F1}
\end{figure*}
We have mentioned before that the total spectroscopic completeness
of our sample is 96\% to $B=20$. Sky regions with the highest
galaxy density, such as groups, might in principle suffer from
higher incompleteness, due to the difficulty to place fibers close
together. This problem is strongly mitigated in the MGC sample,
because it is a combination of three different spectroscopic
campaigns (SDSS, 2dF and MGCz).  However, in order to double-check
the completeness in groups, we performed an additional test.  For
each PM2GC group, within the angular radius corresponding to
$R_{200}$ from the group's centre, we counted the number of MGC
galaxies with redshift and the total number of galaxies in the
photometric catalog brighter than $B=20$.  The ratio of the sum of
all galaxies with redshift and the sum of all galaxies in the
photometric sample for all groups is $r_1$.  Using the same
$R_{200}$ we considered for each group a random RA, DEC for the
group centre within the MGC area and counted again the number of
galaxies with redshift and the number in the photometric catalog
brighter than $B=20$ within the same $R_{200}$. For each group we
repeated this 100 times, finding each time a value $r_2$. The mean
value of the ratio $r_1/r_2$ is 0.999 confirming the high
spectroscopic completeness of the galaxy redshift catalogue even
for galaxies preferentially clustered in groups.

Moreover, we analyzed our 176 groups to assess whether they are
fully contained in the narrow strip of the MGC survey: the aim was
to understand how many and which groups suffer from edge problems
and therefore need to be treated with caution in the subsequent
analysis.  Looking at the group centre position coordinates, we
flagged those 66 groups for which $RA_{centre}\pm R_{200}$ and/or
$DEC_{centre}\pm R_{200}$ fall out of the ranges of the MGC
strip.\footnote{The fractional area lost adopting these limits
instead of $RA_{centre}\pm 1.5 \, R_{200}$ and $DEC_{centre}\pm
1.5 \, R_{200}$ is negligible, therefore also the difference in
number of galaxies falling outside of the field is irrelevant.}
Six of these had clear edge problems also from the comparison with
2PIGG (see \S6).
\begin{figure}
    \centering
    \includegraphics[scale=0.4]{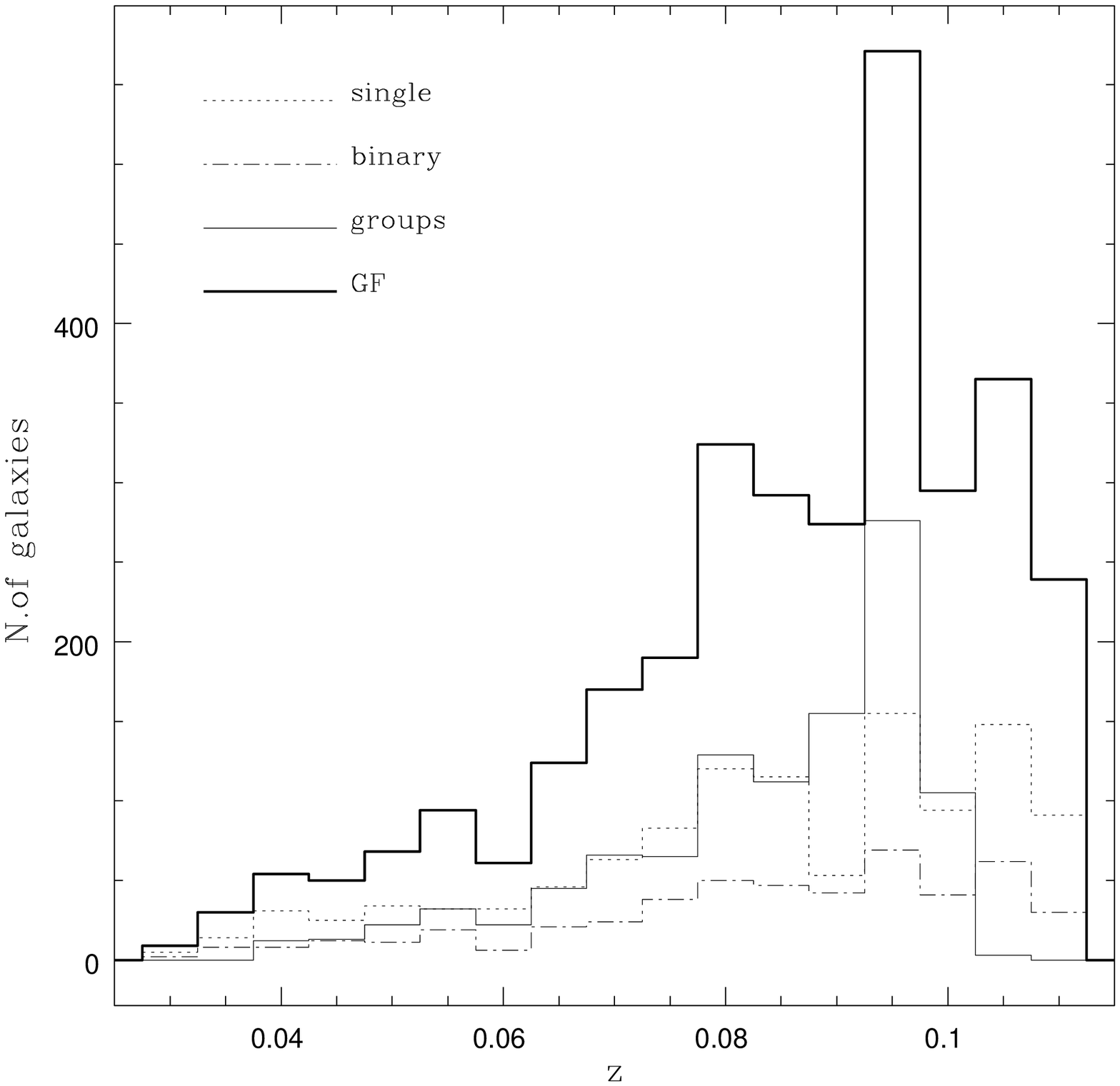}
    \caption{Redshift distribution for the different samples of galaxies: 1141 field-single galaxies (dotted line, 0.03$\leq$z$\leq$0.11), 490 field-binary galaxies (dot-short dashed line, 0.03$\leq$z$\leq$0.11), 1057 group galaxies (solid thin line, 0.04$\leq$z$\leq$0.1) and 3210 general field (GF) galaxies (solid thick line, 0.03$\leq$z$\leq$0.11).}\label{F2}
\end{figure}
\section{Other environments}

Galaxies that are not members of our 176 groups are treated
separately, to study galaxy properties in several environments and
compare the results.

We named ``field-single'' and ``field-binary'' those subsets of
galaxies at 0.03$\leq$z$\leq$0.11 that have no friends (1141) or
solely one friend (490) in their original trial cylindrical
volume, respectively. The first sample, which contains isolated
galaxies, is considered as pure field; the second one is composed
of binary systems of galaxies, i.e. those pairs of bright galaxies
that have a projected mutual separation within 0.5$h^{-1}$Mpc and
a redshift within 1500$\rm km \, s^{-1}$.

The remaining 522 galaxies that are not in groups, field-single or
field-binary environments
are either those that are in groups at z$<$0.04 or z$>$0.1 (302)
or those galaxies that, although located in a trial group, did not
make it into the final group sample (220).
These galaxies are part of the outer regions of groups (outside
$1.5 \, R_{200}$), therefore we prefer not to consider them as
``single''.

Finally, the sample of all galaxies in all environments at
0.03$\leq$z$\leq$0.11
is named ``general field'' (GF from now on),  and is
representative of the general field low-z galaxy population. In
Table \ref{t1} we list the number of galaxies in different
environments.

\begin{figure*}
    \centering
   \vspace{-230pt}
    \includegraphics[width=1\textwidth]{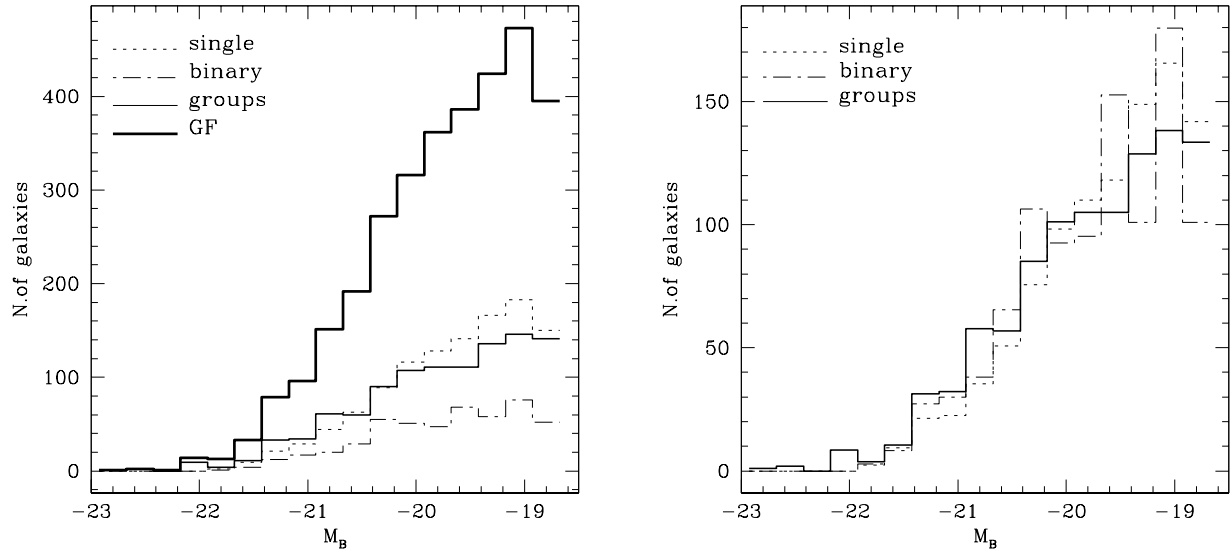}
    \caption{{\bf Left.}  B-band absolute magnitude distribution for the different samples of galaxies: 1141 field-single galaxies (dotted line, 0.03$\leq$z$\leq$0.11), 490 field-binary galaxies (dot-short dashed line, 0.03$\leq$z$\leq$0.11), 1057 group galaxies (solid thin line, 0.04$\leq$z$\leq$0.1) and 3210 general field (GF) galaxies (solid thick line, 0.03$\leq$z$\leq$0.11). {\bf Right.} The absolute magnitude distributions of group, binary and field galaxies, all in the range 0.04$\leq$z$\leq$0.1, normalized to the same total number of galaxies (N=1000).}\label{F3}
\end{figure*}

\section{Properties of groups and galaxies in the different environments}
In Fig.\ref{F1} we show the general characteristics of our group
sample. It is clear that, as in other catalogues, most of the
groups lie in the higher redshift range, and contain fewer than 10
members.

The median redshift and velocity dispersion of the sample are
0.0823 and 191.8 $\rm km \, s^{-1}$, respectively. The range of
velocity dispersion is between 100 $\rm km \, s^{-1}$ and 800 $\rm
km \, s^{-1}$for most groups, with 11 per cent having a velocity
dispersion $<$100 $\rm km \, s^{-1}$ and 29 per cent $>$400$\rm km
\, s^{-1}$. Hence, a significant fraction of the structures we
identify have velocity dispersions higher than 400 $\rm km \,
s^{-1}$, which is the commonly adopted limit between groups and
clusters.
The fraction of groups with less than five members is 63 per cent
and 43 per cent have only three members.

In Fig.\ref{F2} we show a comparison of the redshift distribution
of galaxies in the several environments we have identified. We
note the presence of a prominent peak at z$\sim$0.095 in the
general field distribution, due to groups likely belonging to a
quite populated structure at that redshift.

The magnitude distribution of galaxies in the different
environments is shown in Fig.\ref{F3}. Raw numbers are given in
the left panel, while in the right panel the group, binary and
single galaxy distributions, all in the range
0.04$\leq$z$\leq$0.1, have been normalized to the same number of
galaxies (N=1000) to show the differences. From this figure, the
relative proportion of faint galaxies in the single and binary
fields seems higher than in groups.


\section{Galaxy stellar masses}

We determined the stellar masses for all galaxies in our sample
using the \citet{b41} relation according to which, under the
assumption of a universal IMF, the stellar mass-to-light (M/L)
ratio is strongly correlated with the optical colors of the
integrated stellar populations.
\begin{figure}
    \centering
    \includegraphics[scale=0.4]{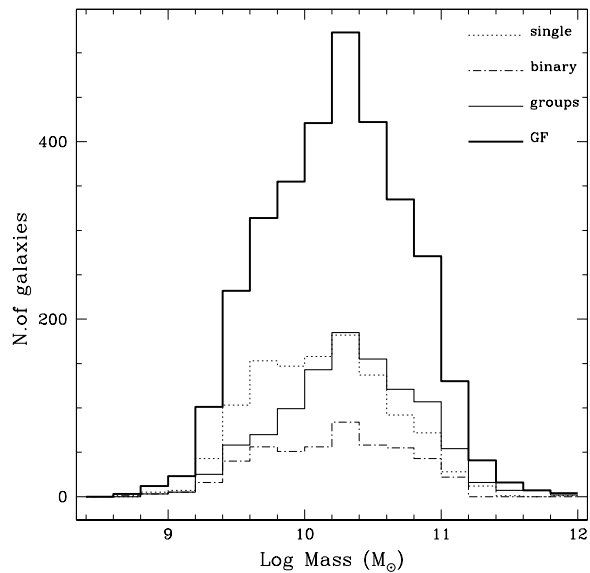}
    \caption{Galaxy stellar mass distribution for the different samples of galaxies: field-single (dot line, 0.03$\leq$z$\leq$0.11), field-binary (dot-short dash line, 0.03$\leq$z$\leq$0.11), groups (solid thin line, 0.04$\leq$z$\leq$0.1) and general field (GF) (solid thick line, 0.03$\leq$z$\leq$0.11) galaxies.}\label{F4}
\end{figure}
Using the B-band photometry, taken from MGCz, we apply the
equation
\begin{equation}
\log_{10}(M_{\star}/L_{B})=a_{B}+b_{B}(B-V) \label{e1}
\end{equation}
having considered a Bruzual \& Charlot model with $a_{B}$=-0.51
and $b_{B}$=1.45 for a \citet{b42} IMF (0.1-125 M$_{\odot}$) and
solar metallicity. To compute the rest frame $(B-V)$ color we
followed the filter conversions from \citet{b43}, i.e.
\begin{equation}
(B-V)_{AB}=0.5928+1.1521[(g-r)-0.6148]\end{equation} using the
MGC-SDSS (from SDSS-EDR/DR1) $(g-r)$ color corrected for
extinction. We then added 0.11 to the $(B-V)$ colors to transform
them from the AB system
 to the Vega system, and
applied the k-corrections in B and V to obtain the rest frame
$(B-V)$ colors. The galaxy stellar masses found with the
eq.(\ref{e1}) were subsequently scaled to a \citet{b44} IMF to
compare with the SDSS, using a conversion factor from Salpeter to
Kroupa of 1/1.55 \citep{cim}.
\begin{figure}
    \centering
    \includegraphics[scale=0.4]{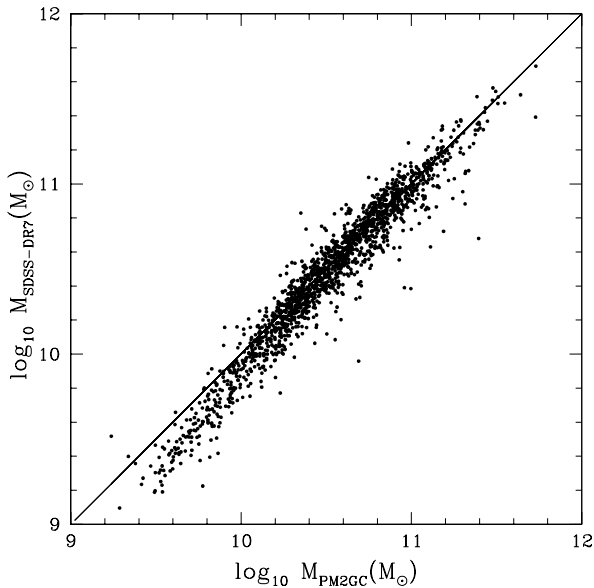}
    \caption{Comparison between the masses of galaxies in our total sample
determined, in this paper, using the \citet{b41} relation and $g$ and $r$ MGC-SDSS magnitudes, and the masses of the same galaxies found in the SDSS-DR7 catalog. The black line is the 1:1 relation.}\label{F5}
\end{figure}
We also took into account the fact that a certain number of
galaxies lie in regions where the photometry can be affected by
CCD edges, satellite trails, bright stars and galaxies
(B$_{MGC}<$12.5), diffraction spikes and so on; any object in
these regions was marked by a flag in the MGCz catalogue to
indicate that it may have an incorrect photometry. Comparing the
B$_{MGC}$ magnitude with the B$_{MGC-SDSS}$ magnitude (determined
using the MGC-SDSS color $g-r$), we have used the B$_{MGC-SDSS}$
magnitude to determine the stellar masses for those galaxies for
which $\Delta B$ = $|$B$_{MGC}$-B$_{MGC-SDSS}$ $|>$ 0.5 mag.

The histogram of the mass distribution for galaxies brighter than
$M_B=$-18.7 in the different environments is shown in
Fig.\ref{F4}.
As discussed in Calvi et al. (2011b) (in preparation), the mass
completeness limit for our sample is $M=10^{10.25} M_{\odot}$, so
any meaningful comparison must be done above this limit. The
variation of the mass function with environment will be discussed
in Calvi et al. (2011b).

We also compared our mass estimates with the stellar masses
computed from the Sloan collaboration from the SDSS-DR7
catalogue\footnote{http://www.mpa-garching.mpg.de/SDSS/DR7/Data/stellarmass.html}
for those galaxies whose MGC and DR7 positions match within 1
arcsec. DR7 masses are computed based on the Sloan photometry,
using model magnitudes, for a Kroupa IMF in the range 0.1-100
$M_{\odot}$ (J. Brinchmann 2010, private communication).

In Fig.\ref{F5} we show this comparison for the GF galaxies in
common with the DR7 mass catalogue.
The agreement is satisfactory at masses above
log$_{10}$(M$_{\star}/$M$_{\odot}$) $\sim$10.3. Here, the
dispersion is similar to the typical mass error for our method
that is normally taken to be 0.2-0.3 dex. At lower masses, there
is a systematic effect, surely due to the different mass estimate
methods, in the sense that our masses are higher than SDSS-DR7
masses by up to $\sim 0.2-0.3$ dex.

Finally, with the \citet{b41} method, we also computed the stellar
masses using the SDSS DR7 model magnitudes corrected for
extinction, obtained from the SDSS Catalogue Archive Server (CAS)
system for 3140 of our galaxies, and verified that these masses
were in very good agreement with those based on MGC-SDSS
magnitudes, without any offset at low masses {(plot not shown)}.

\section{Dataset validation}
The construction of a robust catalogue of groups is essential to
characterise accurately their properties. To validate our
catalogue, we concentrated on a direct comparison with one of the
largest galaxy group sample, the 2PIGG catalogue \citep{b10}. This
consists of $\sim$290000 groups, with at least two members, found
in the Northern and Southern Galactic Patches (NGP and SGP) in the
2dFGRS using a group finding procedure based on a FoF algorithm.
They found galaxy groups using linking parameters calibrated on
realistic mock catalogues identified with high-resolution N-body
simulations and a semi-analytical model of galaxy formation. Their
purpose was to provide overdensity regions that have velocity
dispersions and projected sizes similar to those of their parent
dark matter halos.

The comparison with the 2PIGG catalogue is an important step to
test the work assumptions we made in the FoF algorithm and
validate the effectiveness of our sample. Our galaxy groups are
all contained in their Northern Galactic Patch.

To match our groups with the 2PIGG catalogue, we checked when:
\begin{enumerate}
  \item the geometric centres agreed within 0.1$^{\circ}$; 
  \item the group redshifts differed by at most 0.0007.
\end{enumerate}

\begin{figure*}
    \centering
    \vspace{-240pt}
    \includegraphics[width=1\textwidth]{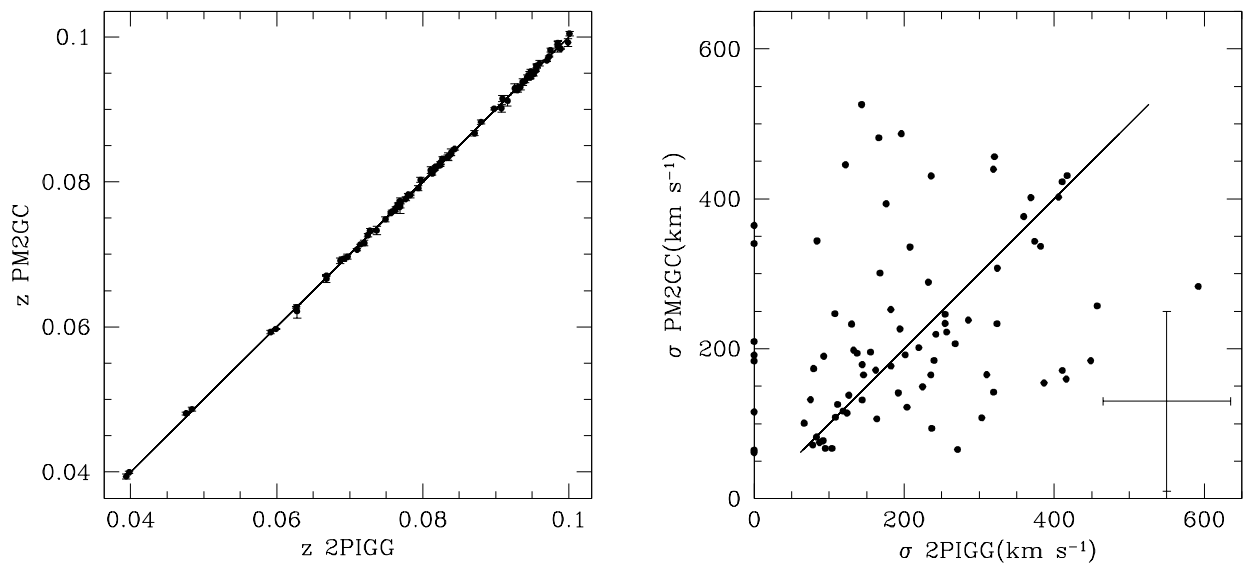}
    \caption{\textbf{Left.} The comparison of redshifts between the 81
    groups matching a 2PIGG group. Our errors are computed by propagating the error
    on the velocities and in many case they are as small as point size. The black line is the 1:1 relation \textbf{Right.} The
    comparison between the velocity dispersions. As the error bars
    are too large and to avoid visual confusion we plotted in the
    bottom right part of the panel the median error of our
    velocity dispersion ($\sim$120 $\rm km \, s^{-1}$) in y and the
    error used by 2PIGG (85 $\rm km \, s^{-1}$) in x. The black line is the 1:1 relation.}
\label{F6}
\end{figure*}

81 of our groups satisfy the above criteria and match a 2PIGG
group. For these, in Fig.\ref{F6} we show the comparison between
our and 2PIGG redshifts (left panel) and velocity dispersions
(right panel). The 2PIGG velocity dispersions used in this plot
are gapper estimates, derived from their tabulated $\sigma$ values
using eq.(4.6) in Eke et al. (2004). 2PIGG adopted a fixed error
on $\sigma$ of 85 $\rm km \, s^{-1}$, which is displayed in
Fig.\ref{F6}. For some 2PIGG groups the sigma value is 0, which
means that the individual galaxy error is at least as big as their
estimate of the velocity dispersion. As expected, given the large
uncertainty in the $\sigma$ measurements based on a few redshifts,
there is a large scatter in this comparison, although 75 per cent
are within the errors.

As a further step, we performed a match between the position of
our group galaxies and that of 2PIGG galaxies within 3 arcsec. All
of these matches agreed also in redshift. This allowed us to
investigate if our group galaxies were associated with the same
2PIGG group matching in geometric centre. In this way we also
found how many of our group galaxies were observed by 2PIGG.

For 23 of our groups none or fewer than 50 per cent of our
galaxies have been observed by 2PIGG. Not surprisingly, these
groups do not match any 2PIGG group according to criteria (1) and
(2) above.

For another 20 of our groups we did not find a match in both
barycentre and redshift with 2PIGG even if at least 50 per cent of
our galaxies have been observed by 2PIGG. However, we have a
higher number of redshifts than 2PIGG in most of these groups.

The remaining 52 of our 176 groups have peculiar characteristics
and deviate from 2PIGG groups. 34 were associated with one group
for 2PIGG but for us they were split in two or more groups; 11
groups have a high velocity dispersion for their number of
galaxies; for the remaining 7 groups, the barycentre of the
corresponding 2PIGG group falls out of the MGC survey strip,
showing that these groups are affected by edge problems in the MGC
thus they will be disregarded in our analysis.
To conclude, about half of the PM2GC groups have a correspondence
in the 2PIGG catalogue. In addition, we found a number of groups
that 2PIGG did not identify. This is due to the higher
spectroscopic completeness of the PM2GC, that contains 1074
galaxies (33 per cent of the PM2GC catalogue) that were not
observed by the 2dFGRS. Moreover, only 20 per cent of our
field-single galaxies belong to a group according to 2PIGG,
confirming the overall statistical agreement with 2PIGG.

\section{PM2GC presentation}
In this last section we describe the five main catalogues we
provide: the group catalogue, and the catalogues of galaxies in
groups, in binary systems, in single field environment and in the
general field.

We stress that the group catalogue contains only groups whose
redshift lies at 0.04$\leq$z$\leq$0.1 because,
 given our selected galaxy redshift range 0.03$\leq$z$\leq$0.11 and
for a maximum velocity dispersion of about 800 $\rm km \, s^{-1}$,
groups at $z$ below $\sim$0.04 and above $\sim$0.1 could be
incomplete. In contrast, the general field, field and binary
catalogues contain all galaxies in the range $0.03\leq z\leq
0.11$.

The main properties of a subsample of galaxy groups in the PM2GC
catalogue are given in Table~2 (corrisponding to Table~2 in
electronic version for the total sample).
  The different columns indicate: (1) PM2GC group serial
number; (2) median group redshift; (3) rest frame velocity
dispersion ($\rm km \, s^{-1}$) computed as in \S3; (4) velocity
dispersion error ($\rm km \, s^{-1}$); (5) number of galaxies
contained within a projected radius $1.5 \, R_{200}$ from the
group geometrical centre and within 3$\sigma$ from the group
redshift; (6-7) geometric centre right ascension and declination
at epoch J2000 (in degrees); (8) $R_{200}$ in Mpc, (9) matched
2PIGG group serial number (9999 if no match with 2PIGG); (10)
2PIGG group, or groups, to which galaxies of our group belong;
(11) number of galaxies of our group that are members of the 2PIGG
matched group (0 indicate no galaxies in 2PIGG or ungrouped
galaxies); (12) 2PIGG match flag (1=good match with 2PIGG, both in
geometric centre and $\Delta z \leq$0.0007, one to one group
correspondance, all of our galaxies in common with 2PIGG are in
the matched 2PIGG group; 2= as 1 but at least one of our group
galaxies is either ungrouped for 2PIGG, and/or belongs to a
different 2PIGG group, and/or is not present in 2PIGG; 3 = no
match with 2PIGG, 2PIGG has none or $<50$ per cent of our
redshifts; 4= no match with 2PIGG, and 2PIGG has at least 50 per
cent of our redshifts; 5= one group for 2PIGG, two or more groups
for us; 6= no match with 2PIGG, high $\sigma$ for number of
members; 0= group with likely edge problems by comparison with
2PIGG), (13) Edge flag, see \S3 1=no edge problem; 2= edge
problem.

In the galaxy catalogues we list the properties of galaxies in
different environments. In the catalogue of galaxies in groups,
see Table~3 for a subsample of galaxies (corrisponding to Table~3
in electronic version for the total catalogue), the columns
indicate: (1) galaxy serial number in MGC; (2) galaxy redshift in
MGC; (3-4) right ascension and declination in MGC at epoch J2000
(in degrees); (5) group number in PM2GC; (6) distance from group
geometric centre in Mpc; (7) distance from group geometric centre
in $R_{200}$ units; (8) logarithm of galaxy stellar mass (Kroupa
IMF) obtained from MGC-SDSS colors (PM2GC mass); (9) logarithm of
galaxy stellar mass (Kroupa IMF) we computed using DR7-CAS colors;
(10) logarithm of galaxy stellar mass (Kroupa IMF) given in the
SDSS-DR7 catalogue (9999 = not available in SDSS-DR7); (11) DR7
$g$ model magnitude corrected for extinction; (12) DR7 $r$ model
magnitude corrected for extinction; (13) DR7 $i$ model magnitude
corrected for extinction; (14) DR7 $u$ model magnitude corrected
for extinction; (15) DR7 $z$ model magnitude corrected for
extinction; (16) rest frame $B-V$ color computed using the
MGC-SDSS $g-r$ color; (17) rest frame $B-V$ color computed using
DR7 $g-r$ color; (18) absolute magnitude in the $B$ band; (19)
absolute magnitude in $V$ band; (20) 2PIGG galaxy serial number
(9999 if not present in 2PIGG); (21) 2PIGG galaxy group (0 if it
is ungrouped, 9999 if galaxy not present in 2PIGG); (22) 2PIGG
match flag of hosting group as in column (12) of Table~2; (23)
edge flag of the hosting group as in column (13) of Table~2.

The catalogues for ''field-single'' and ''field-binary'' galaxies,
see Tables~4 and ~5 for a subsample (corrisponding to Table~4 and
~5 in electronic version for the total catalogues) include the
following columns: (1) galaxy serial number in MGC; (2) galaxy
redshift in MGC; (3-4) right ascension and declination at epoch
J2000 (in degrees); (5) logarithm of galaxy stellar mass (Kroupa
IMF) computed using MGC-SDSS colors; (6) logarithm of galaxy
stellar mass (Kroupa IMF) we computed using DR7-CAS colors, (7)
logarithm of galaxy stellar mass given in the SDSS-DR7 catalogue
(9999 = not available in SDSS-DR7); (8) DR7 $g$ model magnitude
corrected for extinction; (9) DR7 $r$ model magnitude corrected
for extinction; (10) DR7 $i$ model magnitude corrected for
extinction; (11) DR7 $u$ model magnitude corrected for extinction;
(12) DR7 $z$ model magnitude corrected for extinction; (13) rest
frame $B-V$ color computed using the MGC-SDSS $g-r$ color, (14)
rest frame $B-V$ color using the DR7 $g-r$ color; (15) absolute
magnitude in $B$ band; (16) absolute magnitude in $B$ band; (17)
2PIGG galaxy serial number (9999 if not present in 2PIGG), (18)
2PIGG host group, or groups, associated (0 if it is ungrouped,
9999 if galaxy not present in 2PIGG).

Finally, the ``general-field'' catalogue in Table~6 for a
subsample (corrisponding to Table~6 in electronic version for the
total catalogue) has the same entries as the ``field-single'' and
``field-binary'' catalogues, with the addition of a column (19)
listing the environment to which the galaxy belongs (the group
number in PM2GC=group, 1=field-single, 2=field-binary, 3= group at
z$<$0.04 or z$>$0.1, 4= members of trial group, but not in final
group).

\section{Summary and conclusions}

Applying a FoF algorithm to a complete spectroscopic sample of
galaxies brighter than $M_{B}$=-18.7, in the redshift range
0.03$\leq$z$\leq$0.11, taken from MGC survey, we have built the
Padova-Millennium Galaxy and Group Catalogue (PM2GC), a galaxy
sample representative of the general field population in the local
Universe. In the paper we have first described the FoF
group-finding method, calibrated on similar parameters used to
identify groups at high redshift, and then we have presented the
properties of the group and galaxy catalogues obtained.

1057 galaxies
belong to 176 groups containing at least three members at
0.04$\leq$z$\leq$0.1. In addition to the group catalogue
containing the main group properties (redshift, velocity
dispersion, geometric centre etc.), we provide catalogues for
galaxies in groups, galaxies in binary systems and isolated
galaxies. The binary and single catalogues contain 490 and 1141
galaxies, respectively, at 0.03$\leq$z$\leq$0.11.
We also present a general field (GF) catalogue, which comprises
all 3210 galaxies at 0.03$\leq$z$\leq$0.11, representative of the
general field population, including 522 galaxies that either are
in groups at z$<$0.04 or z$>$0.1 or are located in a trial group,
but not in the final group sample.

We have also determined the galaxy stellar masses for all galaxies
in the PM2GC catalogue both using MGC-SDSS magnitudes and SDSS-DR7
CAS magnitudes, and we have showed the existence of a good
agreement with DR7 mass estimates.

In order to validate the effectiveness of our method and the
robustness of our results, we tested our groups comparing them
with the 2PIGG catalogue, finding a good correspondence with 2PIGG
for about half of our groups,
%
and identifying a significant number of groups not present in
2PIGG, thanks to the higher spectroscopic completeness.

The PM2GC provides a valuable database for a different number of
studies which will allow us a better understanding of the
environmental influence on galaxy properties. In addition to
galaxy stellar masses, we have estimated galaxy morphologies,
surface brightness parameters, star formation histories and local
galaxy densities for all PM2GC galaxies that will be presented in
forthcoming papers. In the first upcoming paper (Calvi et al.
2011b), we will analyze the mass functions and the morphological
distributions in different environments comparing the results with
those obtained for WINGS clusters \citep{c8}. Other studies are in
progress, i.e. the analysis of the star formation histories as a
function of environment from spectral information,
the analysis of the galaxy mass-size relation in the local field
and of the variation of the galaxy mass function with local
density.

\section*{Acknowledgments}
We would like to thank the team of the Millennium Galaxy Catalogue
for the excellent dataset they have made available to the
community. Special thanks to Joechen Liske, who provided us the
DVD version of the MGC database and very useful elucidations. We
also would like to thank Vincent Eke for help in using the 2PIGG
catalogue, Jarle Brinchmann for clarifications about the DR7
masses, Antonio Cava, Andrea Biviano and the whole WINGS team for
useful discussions. BV and BMP acknowledge financial support from
ASI Contract I/016/07/0.

\begin{landscape}
\begin{table}
\centering 
\begin{tabular}{ccccccccccccc}
\hline \hline
IDGRPM2 & zPM2 & sigPM2 & ersigPM2 & Ngal & raPM2 &  decPM2 & R200 & IDGR2P & IDGRgal2P & Gal\_ in & flag\_ type & flag\_ bor\_ R\\
            &  & $\rm km \, s^{-1}$ & $\rm km \, s^{-1}$ & &
            deg(J2000) & deg(J2000) & Mpc & & & & & \\
(1) & (2) & (3) & (4) & (5) & (6) & (7) & (8) & (9) & (10) & (11) &
            (12) & (13)\\
\hline \hline
 GR1002 & 0.0769 & 481 & 389 & 6 & 161.6724 & -0.0394 & 1.148 & 2963 & 2963 & 6 & 2 & 1\\
 GR1007 & 0.0944 & 196 & 150 & 6 & 151.4264 & 0.0807 & 0.462 & 4921 & 4921 & 3 & 2 & 1\\
 GR1025 & 0.05 & 88 & 45 & 5 & 172.0366 & 0.1444 & 0.213 & 9999 & 1303 & 3 & 4 & 1\\
 GR1059 & 0.0782 & 222 & 135 & 6 & 182.876 & 0.0627 & 0.529 & 3205 & 3205 & 5 & 2 & 1\\
 GR1065 & 0.0774 & 456 & 239 & 5 & 168.8397 & -0.1213 & 1.088 & 2982 & 2982 & 3 & 2 & 2\\
 GR1072 & 0.0819 & 279 & 273 & 5 & 200.7232 & -0.2424 & 0.664 & 3592 & 3592 & 1 & 5 & 2\\
 GR1100 & 0.0486 & 177 & 223 & 3 & 191.2142 & -0.0579 & 0.427 & 1206 & 1206 & 1 & 2 & 1\\
 GR1123 & 0.0823 & 116 &  85 & 3 & 199.8421 & 0.1619 & 0.275 & 3736 & 3736 & 2 & 2 & 1\\
 GR1135 & 0.1001 & 211 &  88 & 4 & 182.0762 & 0.0777 & 0.498 & 5252 & 5252 & 1 & 4 & 1\\
 GR1181 & 0.0987 & 439 &  68 & 7 & 169.137  & 0.0621 & 1.036 & 5240 & 5240 & 5 & 2 & 1\\
\hline
\end{tabular}
\caption{Subsample of 10 galaxy groups with
their properties. Columns: (1) PM2GC group serial number; (2)
median group redshift; (3) rest frame velocity dispersion; (4)
velocity dispersion error; (5) number of galaxies in group; (6-7)
geometric centre right ascension and declination in degrees; (8)
$R_{200}$ in Mpc, (9) matched 2PIGG group serial number (9999 if
no match with 2PIGG); (10) 2PIGG group, or groups, to which
galaxies of our group belong; (11) number of galaxies of our group
that are members of the 2PIGG matched group (0 indicate no
galaxies in 2PIGG or ungrouped galaxies); (12) 2PIGG match flag
(1=good match with 2PIGG, both in geometric centre and $\Delta z
\leq$0.0007, one to one group correspondance, all of our galaxies
in common with 2PIGG are in the matched 2PIGG group; 2= as 1 but
at least one of our group galaxies is either ungrouped for 2PIGG,
and/or belongs to a different 2PIGG group, and/or is not present
in 2PIGG; 3 = no match with 2PIGG, 2PIGG has none or $<50$ per
cent of our redshifts; 4= no match with 2PIGG, and 2PIGG has at
least 50 per cent of our redshifts; 5= one group for 2PIGG, two or
more groups for us; 6= no match with 2PIGG, high $\sigma$ for
number of members; 0= group with likely edge problems by
comparison with 2PIGG), (13) Edge flag, see \S3 1=no edge problem;
2= edge problem.\label{tg}}
\end{table}
\end{landscape}

\begin{landscape}
\begin{table}
\centering 
\begin{tabular}{ccccccccccccc}
\hline \hline
IDMGC &  zMGC  &   raMGC  &   decMGC  &  IDGRPM2 & d\_ Mpc &  d\_ R200 &  MassPM2 &  MassCAS & MassDR7 &  g\_ cor &   r\_ cor  &  i\_ cor\\
      &        & deg(J2000) & deg(J2000) &       & Mpc & $R_{200}$ &
      $\rm log_{10}M^{\star}$ & $\rm log_{10}M^{\star}$ & $\rm log_{10}M^{\star}$
      &    &     &\\
(1) & (2) & (3) & (4) & (5) & (6) & (7) & (8) & (9) & (10) & (11) &
            (12) & (13)\\
\hline \hline
  1093  &  0.089  &  150.2561  & 0.0216  &  GR1011& 1.162 &  4.834 &   10.9  &    10.83  &   10.97  &   17.431 &   16.472  & 16.014\\
  1241 &   0.0657 &  150.8201 &  -0.0045  & GR1013 &0.256 &  3.044  &  9.98   &   9.96   &   10.04  &   17.863 &  17.253 &  16.909\\
  1371 &  0.0672  & 150.8934 &  0.0761  &  GR1013 &0.244 &  2.831 &   10.95  &   10.92   &  11.0  &   16.827 &  15.846 &  15.373\\
  1418  &  0.0976 &  151.0467 &  -0.164  &  GR1145 &0.184  & 0.753  &  9.11   &   9.08   &   9999.0  &  19.281 &  19.11  &  18.984\\
  1426  &  0.1001 &  151.0606  & -0.1724  & GR1145 &0.086 &  0.348  &  9.45  &    9.42  &    9999.0  &  19.396 &  18.987 &  18.736\\
  1468 &   0.0955 & 151.0964 &  -0.213  &  GR1145 &0.261  & 1.104  &  10.83  &   10.75  &   10.69  &   17.354 &  16.627  & 16.248\\
  1565  &  0.0674  & 150.7827  & 0.0452  &  GR1013 &0.352  & 4.13  &   11.37  &   11.34  &   11.29  &   15.436  & 14.584  & 14.195\\
  1580  &  0.0661 &  150.9395 &  0.03  &    GR1013 & 0.372 & 4.462  &  10.84  &   10.81  &   10.82  &   16.886 & 15.989  & 15.575\\
  1711 &   0.0942  & 151.364 &  0.1374  &  GR1007 &0.529  & 7.548  &  10.26  &   10.23  &   9999.0  &  18.459  & 17.693 &  17.329\\
  1825  &  0.0947  & 151.441  &  0.0789  &  GR1007 &0.092 &  1.333 &   9.5  &     9.39  &   9999.0  &  19.185 &  18.853  & 18.586\\
& & & & & &\rule{0mm}{0.5cm} & & & & & &\\
\hline
& & & & & &\rule{0mm}{0.5cm} & & & & & &\\
\hline \hline
IDMGC & u\_ cor &   z\_ cor &   BV   &    BVcas  &  MabsB  &   MabsV  &   ID2P &   GR2P &  flag\_ type & flag\_ bor\_ R & &\\
 & &  &  &  &  & & & & & & &\\
(1) & (14) & (15) & (16) & (17) & (18) & (19) & (20) & (21) &
            (22) & (23) & & \\
\hline \hline
 1093 & 19.467 &  15.654 &  0.883  &  0.838  &  -20.34  &  -21.223 &  15635 &  4397 &  4   &   2 & & \\
 1241 & 19.276 &  16.645  & 0.55   &  0.535  &  -19.258 &  -19.808  & 15392 &  2219 &  2   &   1  & & \\
 1371 & 18.749 &  15.003 &  0.951 &   0.934  &  -20.217 &  -21.168 &  15368  & 2219 &  2  &    1  & & \\
 1418 & 20.276 &  18.968  & 0.053  &  0.031 &   -18.891  & -18.945  & 9999   & 9999  & 6    &  2  & & \\
 1426 & 20.312 &  18.675 &  0.325  &  0.305  &  -18.744  & -19.069  & 9999  &  9999 &  6  &    2  & &  \\
 1468 & 18.849 &  16.003  & 0.664 &   0.61  &   -20.968  & -21.632 &  15271 &  5069 &  6   &   2  & &  \\
 1565 & 17.332 &  13.923  & 0.812 &   0.785  &  -21.796 & -22.608 &  15410 &  2219 &  2   &   1   & &   \\
 1580 & 18.872 &  15.254 &  0.857  &  0.837  &  -20.294 &  -21.151 &  9999  &  9999  & 2   &   1  & &  \\
 1711 & 19.824 &  17.04  &  0.676  &  0.656  &  -19.498 &  -20.174 &  15183  & 4921 &  2   &   1  & &  \\
 1825 & 20.045 &  18.374 &  0.294  &  0.218  & -18.985 &  -19.279 &  9999  &  9999 &  2   &   1   & &   \\
\hline
\end{tabular} 
\centering \caption{Subsample of
galaxies in groups with their properties. Columns: (1) galaxy
serial number in MGC; (2) galaxy redshift in MGC; (3-4) right
ascension and declination in MGC at epoch J2000 (in degrees); (5)
group number in PM2GC; (6) distance from group geometric centre in
Mpc; (7) distance from group geometric centre in $R_{200}$ units;
(8) logarithm of galaxy stellar mass (Kroupa IMF) obtained from
MGC-SDSS colors (PM2GC mass); (9) logarithm of galaxy stellar mass
(Kroupa IMF) we computed using DR7-CAS colors; (10) logarithm of
galaxy stellar mass (Kroupa IMF) given in the SDSS-DR7 catalogue
(9999 = not available in SDSS-DR7); (11) DR7 $g$ model magnitude
corrected for extinction; (12) DR7 $r$ model magnitude corrected
for extinction; (13) DR7 $i$ model magnitude corrected for
extinction; (14) DR7 $u$ model magnitude corrected for extinction;
(15) DR7 $z$ model magnitude corrected for extinction; (16) rest
frame $B-V$ color computed using the MGC-SDSS $g-r$ color; (17)
rest frame $B-V$ color computed using DR7 $g-r$ color; (18)
absolute magnitude in the $B$ band; (19) absolute magnitude in $V$
band; (20) 2PIGG galaxy serial number (9999 if not present in
2PIGG); (21) 2PIGG galaxy group (0 if it is ungrouped, 9999 if
galaxy not present in 2PIGG); (22) 2PIGG match flag of hosting
group as in column (12) of Table~2; (23) edge flag of the hosting
group as in column (13) of Table~2.\label{tg3}}
\end{table}
\end{landscape}

\begin{landscape}
\begin{table}
\centering 
\begin{tabular}{cccccccccccccccccc}
\hline\hline
IDMGC &  zMGC &    raMGC &     decMGC &   MassPM2 &  MassCAS &  MassDR7 &  g\_ cor &   r\_ cor &    i\_ cor &   u\_ cor  &  z\_ cor &   BV   &    BVcas &   MabsB &    MabsV &    ID2P &   2PGROUP\\
& & deg(J2000) & deg(J2000) & $\rm log_{10}M^{\star}$ &
$\rm log_{10}M^{\star}$ & $\rm log_{10}M^{\star}$ & &  &  & & & & & & & & \\
(1) & (2) & (3) & (4) & (5) & (6) & (7) & (8) & (9) & (10) & (11) &
            (12) & (13)  & (14) & (15) & (16) & (17) & (18)\\
\hline \hline
  19621 &  0.0681  & 170.5799 &  -0.1446 &  10.15  &   10.11  &   10.08  &   17.573 &  17.001 &  16.664 &  18.783 &  16.473  & 0.519  &  0.491  &  -19.798 &  -20.318 &  23969 &  2325\\
  19714 &  0.0689  & 170.5439  & -0.231   & 10.38   &  10.37   &  10.5    &  17.871  & 16.994 &  16.596 &  19.862 &  16.246 &  0.818  &  0.813  &  -19.276  & -20.095 &  23995 &  2325\\
  28996  & 0.0971  & 182.1661  & 0.2117   & 9.85    &  9.83    &  9999.0  &  18.446  & 18.016 &   17.655 &  19.432 &  17.538  & 0.344 &   0.329 &   -19.684 &  -20.028  & 27452 &  5048\\
  29003  & 0.1017  & 182.1215  & 0.2215   & 9.93    &  9.91    &  9999.0   & 19.491  & 18.741 &   18.339 &  20.709 &  18.077  & 0.653 &   0.636 &   -18.762  & -19.415  & 9999   & 9999\\
  9645  &  0.0823  & 159.3158  & 0.0804   & 10.91   &  10.92   &  10.92   &  16.098  & 15.499 &   15.103 &  17.521 &  14.894 &  0.519  &  0.524 &   -21.689 &  -22.207  & 20547 &  3752\\
  9715   & 0.0826  & 159.3208  & 0.0722   & 9.94    &  9.94    &  9999.0  &  18.629  & 18.007 &   17.624 &  19.898 &  17.335 &  0.553 &   0.551 &   -19.15 &   -19.704 &  20544  & 3752\\
  23429  & 0.1064  & 175.4133  & 0.214    & 10.8    &  10.8    &  10.84   &  17.963  & 17.03 &   16.589 &  19.582 &  16.261  & 0.81 &    0.808 &   -20.364 &  -21.174  & 26027  & 0\\
  23444  & 0.1066  & 175.4055  & 0.2344   & 9.67    &  9.66    &  9999.0  &  19.553  & 19.023 &  18.693 &  20.525 &  18.587  & 0.455  &  0.445  &  -18.832 &  -19.288 &  9999  &  9999\\
  96374 &  0.0917  & 161.97    & 0.216    & 10.18   &  10.22   &  10.07   &  17.663  & 17.211 &   16.913 &  18.866 &  16.789 &  0.325  &  0.356  &  -20.559  & -20.885  & 19353 &  4661\\
  11804  & 0.0946  & 161.9251  & 0.2837   & 9.36    &  9.34   &   9999.0   & 19.428  & 19.058 &   18.793 &  20.528 &  18.783 &  0.271 &   0.261 &   -18.706 &  -18.977 &  9999  &  9999\\
\hline
\end{tabular}
\centering \caption{Subsample of galaxies in binary
systems catalogue with their properties. Columns: (1) galaxy
serial number in MGC; (2) galaxy redshift in MGC; (3-4) right
ascension and declination in MGC at epoch J2000 (in degrees); (5)
logarithm of galaxy stellar mass (Kroupa IMF) obtained from
MGC-SDSS colors (PM2GC mass); (6) logarithm of galaxy stellar mass
(Kroupa IMF) we computed using DR7-CAS colors; (7) logarithm of
galaxy stellar mass (Kroupa IMF) given in the SDSS-DR7 catalogue
(9999 = not available in SDSS-DR7); (8) DR7 $g$ model magnitude
corrected for extinction; (9) DR7 $r$ model magnitude corrected
for extinction; (10) DR7 $i$ model magnitude corrected for
extinction; (11) DR7 $u$ model magnitude corrected for extinction;
(12) DR7 $z$ model magnitude corrected for extinction; (13) rest
frame $B-V$ color computed using the MGC-SDSS $g-r$ color; (14)
rest frame $B-V$ color computed using DR7 $g-r$ color; (15)
absolute magnitude in the $B$ band; (16) absolute magnitude in $V$
band; (17) 2PIGG galaxy serial number (9999 if not present in
2PIGG); (18) 2PIGG galaxy group (0 if it is ungrouped, 9999 if
galaxy not present in 2PIGG).\label{tg5}}
\end{table}
\end{landscape}

\begin{landscape}
\begin{table}
\begin{tabular}{cccccccccccccccccc}
\hline \hline
IDMGC &  zMGC &    raMGC &     decMGC &   MassPM2 &  MassCAS &  MassDR7 &  g\_ cor &   r\_ cor &    i\_ cor &   u\_ cor  &  z\_ cor &   BV   &    BVcas &   MabsB &    MabsV &    ID2P &   2PGROUP\\
& & deg(J2000) & deg(J2000) & $\rm log_{10}M^{\star}$ &
$\rm log_{10}M^{\star}$ & $\rm log_{10}M^{\star}$ & &  &  & & & & & & & & \\
(1) & (2) & (3) & (4) & (5) & (6) & (7) & (8) & (9) & (10) & (11) &
            (12) & (13)  & (14) & (15) & (16) & (17) & (18)\\
\hline \hline
  61514 &  0.0844 &  215.9539 &   -0.2483 &  10.57  &   10.5  &    10.53 &    17.277 &  16.636 &   16.252 &  18.52  &  16.054 &  0.559  &  0.511 &   -20.708 &   -21.267 &  38599 &  0\\
  27280 &  0.0975 &  179.8905 &  -0.0265 &  10.2  &    10.2 &     9999.0 &   18.757 &  18.033 &   17.606 &  20.054 &   17.385 &  0.612 &   0.607 &   -19.591 &  -20.203 &  27753 &  0\\
  11383  & 0.0635 &   161.5143 &   0.1471 &   10.11 &    10.05 &    9.92  &    16.948 &  16.566 &  16.294 &  18.195 &   16.202 &  0.364  &  0.321 &   -20.256 &  -20.62 &   19581 &  0\\
  66778 &  0.1079 &  220.1103 &  0.0101 &   9.91  &    9.88   &   9999.0 &   18.963 &  18.46 &   18.168  & 20.029 &  18.08 &   0.436 &   0.414  &  -19.487 &  -19.923  & 39496  & 5964\\
  52268 &  0.1076 &  207.9901 &   -0.1881 &  10.47 &    10.43 &    9999.0 &   19.075 &  18.14 &   17.61 &   20.867 &  17.26 &   0.838 &   0.811 &   -19.431 &  -20.27 &   9999 &   9999\\
  56234 &  0.1053 &  211.4459 &  0.0196 &   10.13 &    10.09  &   9999.0 &   18.758 &  18.115 &  17.788  & 20.045 &  17.576 &  0.543 &   0.513 &   -19.668 &  -20.211 &  37628 &  0\\
  16076  & 0.0813 &  167.1715 &  0.1933 &   10.56  &   10.44 &    10.43  &   17.105 &  16.574 &  16.229 &  18.357  & 15.989  & 0.528 &   0.447 &   -20.778 &  -21.307 &  21884 &  0\\
  13606 &  0.108 &   164.0227 &  -0.0388 &  9.95 &     9.88  &    9.74 &     17.877 &  17.636 &   17.419 &  18.692 &  17.365 &  0.158 &   0.112 &   -20.595 &  -20.753 &  22323 &  6222\\
  57725 &  0.0804 &  213.0875 &  -0.1577 &   9.8 &      9.75 &     9.75 &     17.831 &   17.475 &  17.193 &  18.914 &  17.1  &   0.28  &   0.244  &  -19.793 & 20.073  & 37292  & 0\\
  25661 &  0.0594 &  178.081 &   -0.1864 &  10.21 &    10.13  &   10.05  &   17.165 &  16.64 &   16.34  &  18.405 &  16.131  & 0.549 &   0.495  &  -19.838 &  -20.387 &  25136 &  0\\ \hline
\end{tabular}
\centering \caption{Subsample of galaxies in single
catalogue with their properties. Columns: (1) galaxy serial number
in MGC; (2) galaxy redshift in MGC; (3-4) right ascension and
declination in MGC at epoch J2000 (in degrees); (5) logarithm of
galaxy stellar mass (Kroupa IMF) obtained from MGC-SDSS colors
(PM2GC mass); (6) logarithm of galaxy stellar mass (Kroupa IMF) we
computed using DR7-CAS colors; (7) logarithm of galaxy stellar
mass (Kroupa IMF) given in the SDSS-DR7 catalogue (9999 = not
available in SDSS-DR7); (8) DR7 $g$ model magnitude corrected for
extinction; (9) DR7 $r$ model magnitude corrected for extinction;
(10) DR7 $i$ model magnitude corrected for extinction; (11) DR7
$u$ model magnitude corrected for extinction; (12) DR7 $z$ model
magnitude corrected for extinction; (13) rest frame $B-V$ color
computed using the MGC-SDSS $g-r$ color; (14) rest frame $B-V$
color computed using DR7 $g-r$ color; (15) absolute magnitude in
the $B$ band; (16) absolute magnitude in $V$ band; (17) 2PIGG
galaxy serial number (9999 if not present in 2PIGG); (18) 2PIGG
galaxy group (0 if it is ungrouped, 9999 if galaxy not present in
2PIGG).\label{tg4}}
\end{table}
\end{landscape}

\begin{landscape}
\begin{table}
\begin{tabular}{ccccccccccccccccccc}
\hline \hline
IDMGC  & zMGC  &   raMGC    &  decMGC  &  MassPM2 &  MassCAS &  MassDR7 &  g\_cor &   r\_cor &   i\_cor &   u\_cor &   z\_cor  &  BV    &   BVcas &   MabsB  &   MabsV &    ID2P &   GR2P &  flag\_env\\
& & deg(J2000) & deg(J2000) & $\rm log_{10}M^{\star}$ &
$\rm log_{10}M^{\star}$ & $\rm log_{10}M^{\star}$ & &  &  & & & & & & & & & \\
(1) & (2) & (3) & (4) & (5) & (6) & (7) & (8) & (9) & (10) & (11) &
            (12) & (13)  & (14) & (15) & (16) & (17) & (18) & (19)\\
\hline \hline
  61514 &  0.0844  & 215.9539  & -0.2483  & 10.57  &   10.5   &   10.53  &   17.277 &  16.636 &  16.252  & 18.52  &  16.054 &  0.559  &  0.511 &   -20.708  & -21.267 &  38599  & 0  &    1\\
  27280 &  0.0975 &  179.8905 &  -0.0265  & 10.2  &    10.2  &    9999.0  &  18.757 &  18.033 &  17.606 &  20.054  & 17.385  & 0.612  &  0.607  &  -19.591 &  -20.203 &  27753  & 0  &    1\\
  11383 &  0.0635  & 161.5143 &  0.1471  &  10.11  &   10.05  &   9.92  &    16.948 &  16.566 &  16.294  & 18.195  & 16.202  & 0.364  &  0.321  &  -20.256  & -20.62  &  19581 &  0  &    1\\
  66778 &  0.1079 &  220.1103  & 0.0101  &  9.91  &    9.88   &   9999.0  &  18.963 &  18.46 &   18.168 &  20.029 &  18.08 &   0.436  &  0.414 &   -19.487 &  -19.923 &  39496 &  5964 &  1\\
  52268 &  0.1076 &  207.9901 &  -0.1881 &  10.47  &   10.43  &   9999.0  &  19.075  & 18.14 &   17.61  &  20.867 &  17.26 &   0.838  &  0.811  &  -19.431  & -20.27 &   9999  &  9999 &  1\\
  56234 &  0.1053 &  211.4459 &  0.0196  &  10.13  &   10.09  &   9999.0  &  18.758 &  18.115 &  17.788 &  20.045 &  17.576 &  0.543 &   0.513  &  -19.668 &  -20.211 &  37628 &  0   &   1\\
  16076 &  0.0813 &  167.1715 &  0.1933  &  10.56  &   10.44  &   10.43  &   17.105 &  16.574  & 16.229  & 18.357 &  15.989 &  0.528 &   0.447  &  -20.778 &  -21.307  & 21884 &  0   &   1\\
  13606 &  0.108 &   164.0227  & -0.0388  & 9.95   &   9.88   &   9.74 &     17.877 &  17.636 &  17.419 &  18.692 &  17.365 &  0.158  &  0.112  &  -20.595 &  -20.753 &  22323 &  6222  & 1\\
  57725 &  0.0804  & 213.0875 &  -0.1577  & 9.8   &    9.75  &    9.75  &    17.831 &  17.475 &  17.193 &  18.914 &  17.1 &    0.28  &   0.244 &   -19.793 &  -20.073 &  37292 &  0  &    1 \\
  25661 &  0.0594 &  178.081 &   -0.1864 &  10.21  &   10.13  &   10.05  &   17.165 &  16.64 &   16.34  &  18.405 &  16.131 &  0.549  &  0.495  &  -19.838 &  -20.387  & 25136 &  0  &    1\\
  60827 &  0.0738 &  215.59   &  -0.1769 &  9.7   &    9.68   &   9.61  &    17.98 &   17.647 &  17.405 &  18.969 &  17.197 &  0.282 &   0.265  &  -19.534  & -19.816  & 38626 &  0  &   1\\
  46480 &  0.0978 &  201.6882 &  -0.1698 &  9.26    &  9.22   &   9999.0  &  19.369  & 19.127 &  18.914 &  20.648 &  18.85 &   0.146 &   0.113  &  -18.933  & -19.079 &  9999  &  9999  & 1 \\
  25285  & 0.0956  & 177.4551 &  -0.2612 &  9.6   &    9.49  &    9999.0  &  19.203 &  18.842  & 18.555 &  20.306 &  18.557 &  0.326 &   0.25   &  -19.115 &  -19.441 &  9999 &   9999  & 1\\
\hline
\end{tabular}
\centering \caption{Subsample of galaxies in general
field catalogue with their properties. Columns: (1) galaxy serial
number in MGC; (2) galaxy redshift in MGC; (3-4) right ascension
and declination in MGC at epoch J2000 (in degrees); (5) logarithm
of galaxy stellar mass (Kroupa IMF) obtained from MGC-SDSS colors
(PM2GC mass); (6) logarithm of galaxy stellar mass (Kroupa IMF) we
computed using DR7-CAS colors; (7) logarithm of galaxy stellar
mass (Kroupa IMF) given in the SDSS-DR7 catalogue (9999 = not
available in SDSS-DR7); (8) DR7 $g$ model magnitude corrected for
extinction; (9) DR7 $r$ model magnitude corrected for extinction;
(10) DR7 $i$ model magnitude corrected for extinction; (11) DR7
$u$ model magnitude corrected for extinction; (12) DR7 $z$ model
magnitude corrected for extinction; (13) rest frame $B-V$ color
computed using the MGC-SDSS $g-r$ color; (14) rest frame $B-V$
color computed using DR7 $g-r$ color; (15) absolute magnitude in
the $B$ band; (16) absolute magnitude in $V$ band; (17) 2PIGG
galaxy serial number (9999 if not present in 2PIGG); (18) 2PIGG
galaxy group (0 if it is ungrouped, 9999 if galaxy not present in
2PIGG); (19) environment to which the galaxy belongs (the group
number in PM2GC=group, 1=field-single, 2=field-binary, 3= group at
z$<$0.04 or z$>$0.1, 4= members of trial group, but not in final
group).\label{tg6}}
\end{table}
\end{landscape}


\label{lastpage}

\end{document}